\RequirePackage{fix-cm}
\documentclass[smallextended]{svjour3}       
\smartqed  
\usepackage{graphicx}
\usepackage{multirow}
\usepackage{graphicx,subfigure}
\usepackage{subcaption}
\usepackage{subfigure}
\begin{document}

\title{Efficient and Accurate Tuberculosis Diagnosis: Attention Residual U-Net and Vision Transformer Based Detection Framework}


\author{Greeshma K.*       \and
        Vishnukumar S. 
}


\institute{Greeshma K. \at
              Department of Computer Applications, Cochin University of Science \& Technology, Kochi - 682 022, Kerala, India \\
              \email{*greeshmavarier@gmail.com}            \\
           \and
           Greeshma K. \at
              Department of Computer Science, Union Christian College, Aluva, Kochi - 683 102, Kerala, India \\
              \and
           Vishnukumar S. \at
              Department of Computer Applications, Cochin University of Science \& Technology, Kochi - 682 022, Kerala, India
}

\date{Received: date / Accepted: date}

\maketitle

\begin{abstract}
Tuberculosis (TB), an infectious disease caused by Mycobacterium tuberculosis, continues to be a major global health threat despite being preventable and curable. This burden is particularly high in low and middle income countries. Microscopy remains essential for diagnosing TB by enabling direct visualization of Mycobacterium tuberculosis in sputum smear samples, offering a cost-effective approach for early detection and effective treatment. Given the labour-intensive nature of microscopy, automating the detection of bacilli in microscopic images is crucial to improve both the expediency and reliability of TB diagnosis. The current methodologies for detecting tuberculosis bacilli in bright field microscopic sputum smear images are hindered by limited automation capabilities, inconsistent segmentation quality, and constrained classification precision. This paper proposes a two-stage deep learning methodology for tuberculosis bacilli detection, comprising bacilli segmentation followed by classification. In the initial phase, an advanced U-Net model employing attention blocks and residual connections is proposed to segment microscopic sputum smear images, enabling the extraction of Regions of Interest (ROIs). The extracted ROIs are then classified using a Vision Transformer, which we specifically customized as TBViT to enhance the precise detection of bacilli within the images.  For the experiments, a newly developed dataset of microscopic sputum smear images derived from g Ziehl-Neelsen-stained slides is used in conjunction with existing public datasets. The qualitative and quantitative evaluation of the experiments using various metrics demonstrates that the proposed model achieves significantly improved segmentation performance, higher classification accuracy, and a greater level of automation, surpassing existing methods.
\end{abstract}

\section{Introduction}
Mycobacterium tuberculosis is the pathogen responsible for tuberculosis (TB), a contagious disease primarily impacting the lungs but capable of affecting other organs like the brain, kidneys, and spine. Untreated TB continues to pose severe health risks and remains life-threatening. According to the 2024 Global Tuberculosis Report by the World Health Organization (WHO), tuberculosis likely regained its status as the top cause of death worldwide from a single infectious agent in 2023 \cite{whotbr2024}. This disease reportedly caused nearly double the fatalities of HIV/AIDS and surpassed COVID-19 in mortality, with an estimated 1.25 million deaths attributed to tuberculosis in 2023. Microscopic examinations for tuberculosis (TB) detection offer several advantages, including affordability, non-invasiveness, accessibility, rapid results, on-site diagnostic capabilities, and the ability to be repeated for accurate evaluations.

\par 
Two microscopy techniques, Fluorescent and Bright Field, are employed in the detection of TB. For Bright Field microscopy, the Ziehl-Neelsen (ZN) staining method is utilized to prepare samples. Initially, sputum samples collected from patients are spread into thin smears on glass slides, which are then fixed and stained with Cabrol fuchsin. When examined under a microscope, acid-fast bacilli appear red or pink against a blue or green background, depending on the counterstain applied after the slide has been decolorized with acid-alcohol. In contrast, fluorescent microscopy requires sputum samples to be stained with Auramine-Rhodamine. These samples are prepared on glass slides, with the smear first stained with Auramine-O before being rinsed to eliminate any excess stain. After applying a decolorizing agent, such as acid-alcohol or acidified ethanol, non-acid-fast organisms and any background staining are effectively removed. Following this, Rhodamine stain is introduced to highlight the non-acid-fast bacteria and background material. In the final stained slide, acid-fast bacilli will appear as green rods set against a dark, almost black, background. Fluorescent microscopy, though offering about 10\% greater sensitivity compared to bright field microscopy \cite{steingart2006fluorescence}, tends to be more expensive. However, bright field microscopy remains a popular choice due to its lower cost, making it accessible for routine diagnostic use. The 2024 WHO Global Tuberculosis Report indicates a rise in global TB cases from 10.0 million in 2020 to 10.3 million in 2021, 10.6 million in 2022, and 10.8 million in 2023 \cite{whotbr2024} \cite{whotbr2023} \cite{whotbr2022}, likely reestablishing TB as the world’s leading infectious disease killer. COVID-19 disruptions have caused delays in TB diagnosis and treatment, intensifying pressures on healthcare systems. Sputum smear microscopy remains essential, confirming 62\% of TB cases bacteriologically in 2023, compared to 48\% confirmed through rapid diagnostic tests \cite{whotbr2023}, and automated microscopy could further boost diagnostic efficiency. Implementing automated systems for TB bacilli detection in microscopic images could therefore greatly alleviate this strain, improving diagnostic efficiency and accuracy and ultimately supporting more effective TB control.
\par
Current methods in this domain demand advancements in automation to achieve more efficient and streamlined workflows. Moreover, enhancing segmentation precision and classification accuracy is critical to improving the overall performance of these techniques. This paper introduces an advanced approach for detecting tuberculosis bacilli, leveraging deep learning techniques for segmentation, and utilizing a powerful model designed to capture fine details in complex image regions. By integrating attention mechanisms with cutting-edge neural networks, the method enhances both segmentation accuracy and focus on relevant regions. Additionally, a custom vision transformer is employed for robust classification, ensuring high precision in identifying bacilli. The structure of the paper is as follows: Section 2 reviews the Related Works, Section 3 outlines the Methodology, Section 4 presents the Experimental Results and Analysis, and Section 5 concludes the study.
\section{Related Works}
In recent years, the focus on automating bacilli detection has shifted towards integrating more advanced image processing techniques and machine learning models. However, the earliest attempts in the late 2000s were centered on basic image segmentation methods aimed at distinguishing bacilli from the background. These early efforts relied on hand-crafted features to segment and quantify bacterial regions. For example, in \cite{costa2008automatic}, a thresholding approach was used based on pixel intensity histograms, followed by size and morphological filters to eliminate irrelevant artifacts. This method achieved a sensitivity of 76.65\% and a false positive rate of 12\%.Later works, such as in \cite{sadaphal2008image}, introduced the use of shape and size parameters like axis length ratio and eccentricity to refine bacilli identification after Bayesian segmentation, although quantification was limited. More recent studies, like \cite{sotaquira2009detection}, have focused on improving both detection and quantification, with techniques like combining results from multiple color spaces (Lab and YCbCr) through logical operations, achieving higher sensitivity (90.9\%) and specificity (100\%). A method proposed in \cite{priya2015separation} addressed the challenge of overlapping bacilli by using the method of concavity (MOC), which outperformed traditional approaches such as marker-controlled watershed (MCW) and multi-phase active contour (MAC) in accurately separating individual bacilli.

\par 
Machine learning methods for detecting bacteria in images started around 2009, developing alongside traditional image processing techniques. Bacilli detection typically follows a two-step process: segmentation to identify the region of interest (ROI), followed by classification into bacilli and non-bacilli categories, which reduces computational demands by focusing only on relevant areas. R. Khutlang et al. automated this process using two one-class classifiers—one for segmentation and the other for classifying the final regions \cite{khutlang2010automated}. Another method \cite{khutlang2009classification} used three pixel classifiers, achieving 88.38\% correctly categorized pixels during segmentation, and tested five classifiers for classification, all showing over 95\% accuracy, sensitivity, and specificity using Fisher-mapped features. A two-stage segmentation approach, combining HSV and CIE Lab* color spaces, was proposed in \cite{zhai2010automatic}, with classification based on three shape feature descriptors: area, compactness, and roughness, using a decision tree. Costa Filho et al. \cite{costafilho2012mycobacterium} introduced a feedforward neural network for segmentation and introduced the color ratio feature, used for classification. In \cite{costa2015automatic}, several handcrafted features were tested for segmentation, with R-G features performing best. In post-processing, SVM \cite {hearst1998support} and feedforward neural networks were employed, with SVM showing the highest sensitivity at 96.8\%. A random forest-based approach\cite{breiman2001random} improved both segmentation and classification in \cite{ayas2014random}. A Gaussian Mixture Model for segmentation, combined with a bacilli counting algorithm \cite{fandriyanto2021detecting}, achieved 93.52\% accuracy using images from the ZNSM-iDB database \cite{shah2017ziehl}.

\par 
Convolutional Neural Networks (CNNs) are commonly used for bacilli detection in bright-field microscopic sputum smear images. Yadini Pérez López et al. \cite{lopez2017automatic} developed a model with three convolutional layers using the image’s R-G feature, achieving a 99\% area under the ROC curve. Another study \cite{kant2018towards} used a deep neural network for TB detection, focusing on bacilli locations, with an emphasis on precision (67.55\%) and recall (83.78\%). Panicker et al. \cite{panicker2018automatic} proposed a CNN with Otsu-based segmentation and connected component analysis for binary classification, yielding 97.13\% recall, 78.4\% precision, and an F-score of 86.76\%. El-Melegy \cite{el2019identification} applied Faster RCNN, followed by a CNN to reduce false positives, achieving 98.4\% recall, 85.1\% precision, and a 91.2\% F-score. Dinesh Jackson Samuel et al. \cite{dinesh2019tuberculosis} used Inception V3 with SVM, reporting 95.05\% accuracy. Another approach \cite{yang2020cnn} combined two CNNs and Logistic Regression, with 87.13\% sensitivity, 87.62\% specificity, and an 80.18\% F1 score. Serrao et al. \cite{serrao2020automatic} introduced mosaic images for bacilli detection, achieving 99\% accuracy, though it is less practical for real-world use. Panicker et al. \cite{panicker2021lightweight} later proposed a lightweight CNN with 97.83\% accuracy, while their enhanced model using DenseNet-121 \cite{panicker2022automatic} improved classification accuracy to 99.7\%. In \cite{greeshma2023identification}, a U-Net model segmented images, followed by Random Forest classification, yielding 93.98\% accuracy. Recently, K.S. Mithra \cite{mithra2023enhanced} used Otsu for segmentation and an enhanced Fuzzy Gaussian Network for classification, with 92.4\% segmentation accuracy and 0.004\% MSE.

\par 

Most works directly input Regions of Interest (ROIs) into models without effective segmentation. Though \cite{mithra2023enhanced} employed Otsu, it struggled with adequate segmentation. Efficient segmentation is crucial for automating bacilli detection, as it reduces classifier burden by narrowing the dataset to smaller ROIs. This work proposes an Attention Residual U-Net for ROI extraction, paired with a visual transformer to enhance accuracy and performance. Existing studies rely on public datasets \cite{ya12-j913-22} \cite{shah2017ziehl}, which may limit generalization. To address this, we created a new dataset, 'DCA-CUSAT Bright Field Microscopic Sputum Smear TB Dataset,' containing 101 images from TB-positive ZN-stained slides from Government District TB Hospital, Ernakulam, Kerala, India.

\section{Methodology}
\begin{figure}[htp]
    \centering
    \includegraphics[scale=0.1]{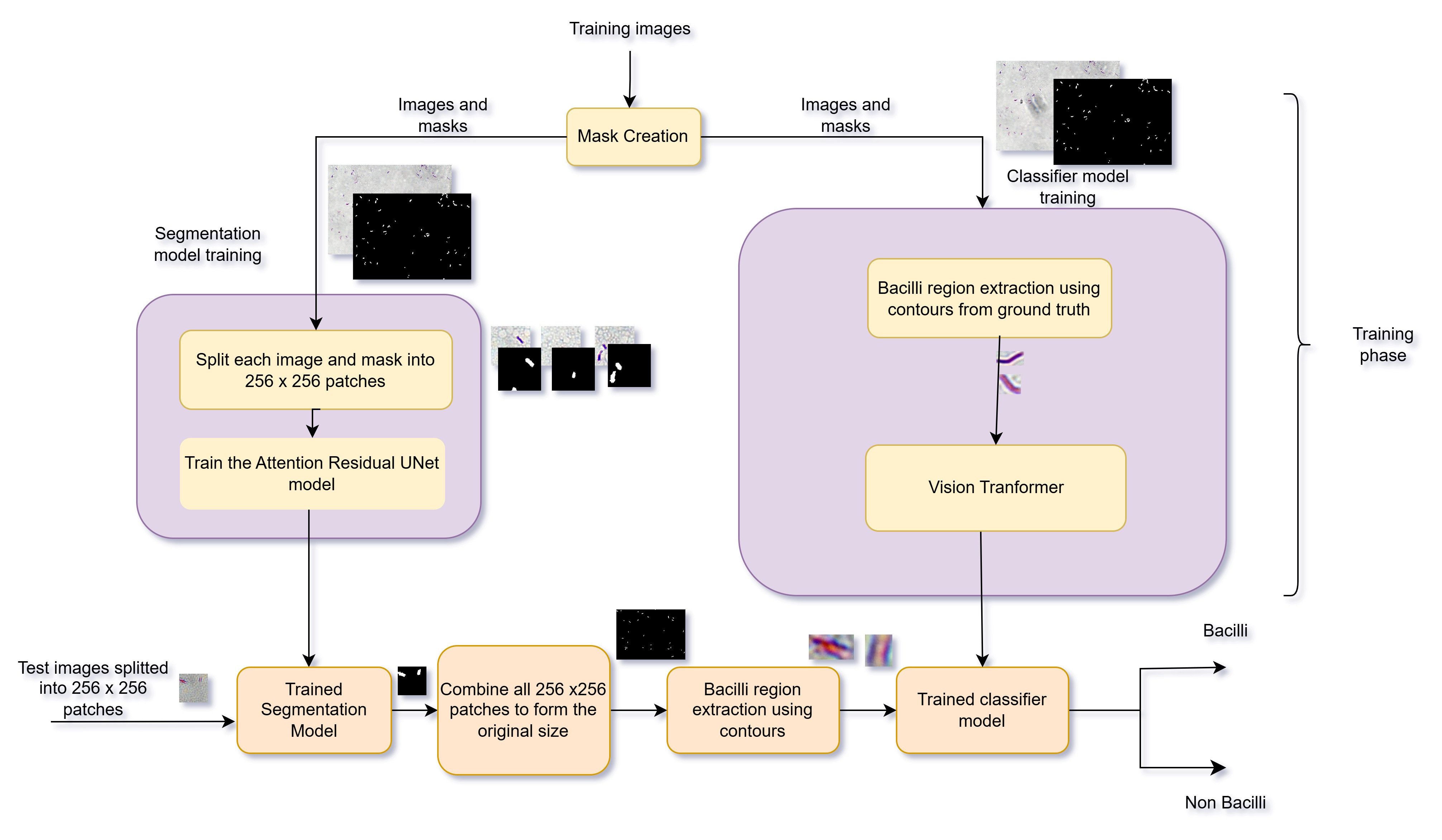}
        \caption{Proposed System Architecture.}
    \label{oas}
\end{figure}

The proposed methodology consists of two phases, segmentation followed by classification. Here we propose a composite U-Net architecture, incorporating attention blocks, \cite{oktay2018attention} and residual connections \cite{alom2018recurrent} for segmentation. This design incorporates residual connections and attention mechanisms to effectively capture multi-scale features and enhance spatial details, employing attention gates to selectively emphasize relevant regions during the upsampling process, ultimately yielding more accurate segmentation results. For classification, a customized Vision Transformer leveraging components such as patch embedding, multi-head attention, and dense layers is employed to distinguish between bacilli and non-bacilli in images. For the training and evaluation of the segmentation and classification models, ground truth binary images were generated using QuPath \cite{bankhead2017qupath}. The overall architecture is depicted in Figure \ref{oas}.

\begin{figure}[htp]
    \centering
    \includegraphics[scale=0.5]{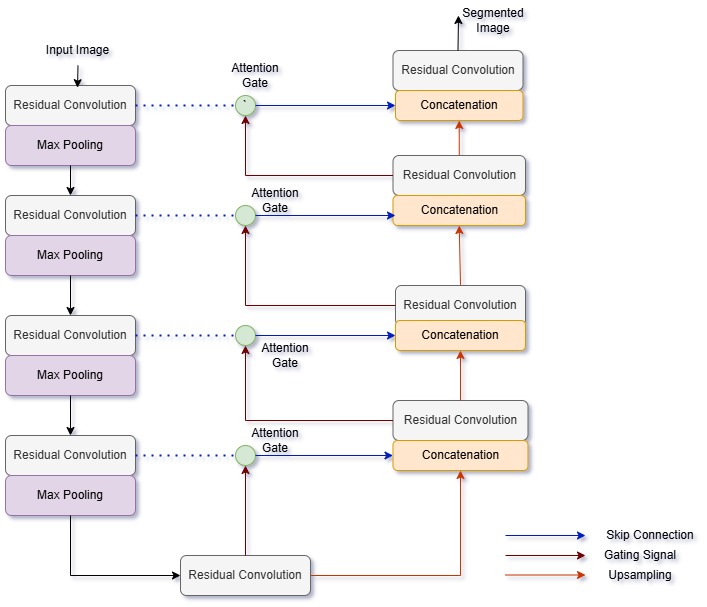}
    \caption{Proposed Attention Residual U-Net for segmentation.}
    \label{pau}
\end{figure}

\par
\begin{figure}[htp]
    \centering
    \includegraphics[scale=0.4]{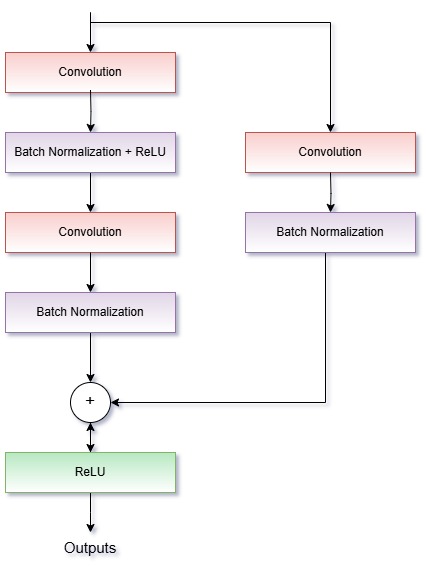}
    \caption{Architecture of the Residual Convolution.}
    \label{arc}
\end{figure}

 
In the proposed segmentation model, residual connections are added in the encoder and decoder paths, and attention gates are integrated into the skip connections. The integration of residual connections to the U-Net architecture, improves model depth and stability, addressing the vanishing gradient problem and enhancing feature accumulation across layers. Meanwhile, attention mechanisms allow the model to prioritize salient features, enabling focused segmentation in heterogeneous or noisy environments. The architecture of the proposed Attention Residual U-Net  is  shown in Figure \ref{pau}. In the downsampling path, the network progressively reduces the spatial dimensions while increasing feature depth through residual convolutional blocks and max-pooling. In the upsampling path, the attention blocks refine the feature maps by concentrating on relevant areas before concatenation with corresponding downsampling features. The model uses a final 1$\times 1$ convolution followed by a sigmoid activation for binary segmentation. Batch normalization is applied at various stages to stabilize training, and dropout is included to prevent overfitting. Each residual block comprises two convolutional layers followed by batch normalization and ReLU activation. The architecture of the residual convolution is depicted in Figure \ref{arc}.

\par 
For efficient segmentation, each image is divided into equal-sized patches, which are segmented individually by the proposed composite Attention Residual U-Net. After segmentation, the patches are reassembled to form a complete mask. In the post-processing phase, contour analysis is applied to this mask to enhance segmentation quality by removing small, extraneous regions. Specifically, contours below a designated area threshold are discarded to reduce noise and exclude non-target segments. Bounding boxes are then generated around the remaining segmented regions, and ROIs are extracted from the corresponding RGB image using the bounding box positions, to isolate each relevant area for further classification.

To effectively perform the classification task, we customized the Vision Transformer (ViT) model, which we term TBViT, specifically adapted for the detection of bacilli. It is specifically adapted to handle the challenges posed by binary classification of ROIs, where high testing imbalance arises due to the segmentation model’s success in isolating bacilli regions. The segmentation model efficiently identifies and extracts bacilli-containing regions, often resulting in a smaller number of non-bacilli areas in the test set. This causes a notable class imbalance in the testing dataset, despite using a balanced dataset for training.

\par
To address this imbalance effectively, the TBViT model incorporates focal loss and adaptive class weighting, which direct the model to focus more on the minority class during testing. Focal loss is particularly effective for emphasizing harder-to-classify samples and compensates for the inherent imbalance by assigning higher weights to underrepresented regions. This setup ensures that the model remains sensitive to non-bacilli areas, even when they appear less frequently in the test data. Class weighting further refines this by dynamically adjusting the model’s attention to class distribution, ensuring robust performance across varying degrees of imbalance.

\begin{figure}[htp]
    \centering
    \includegraphics[scale=0.1]{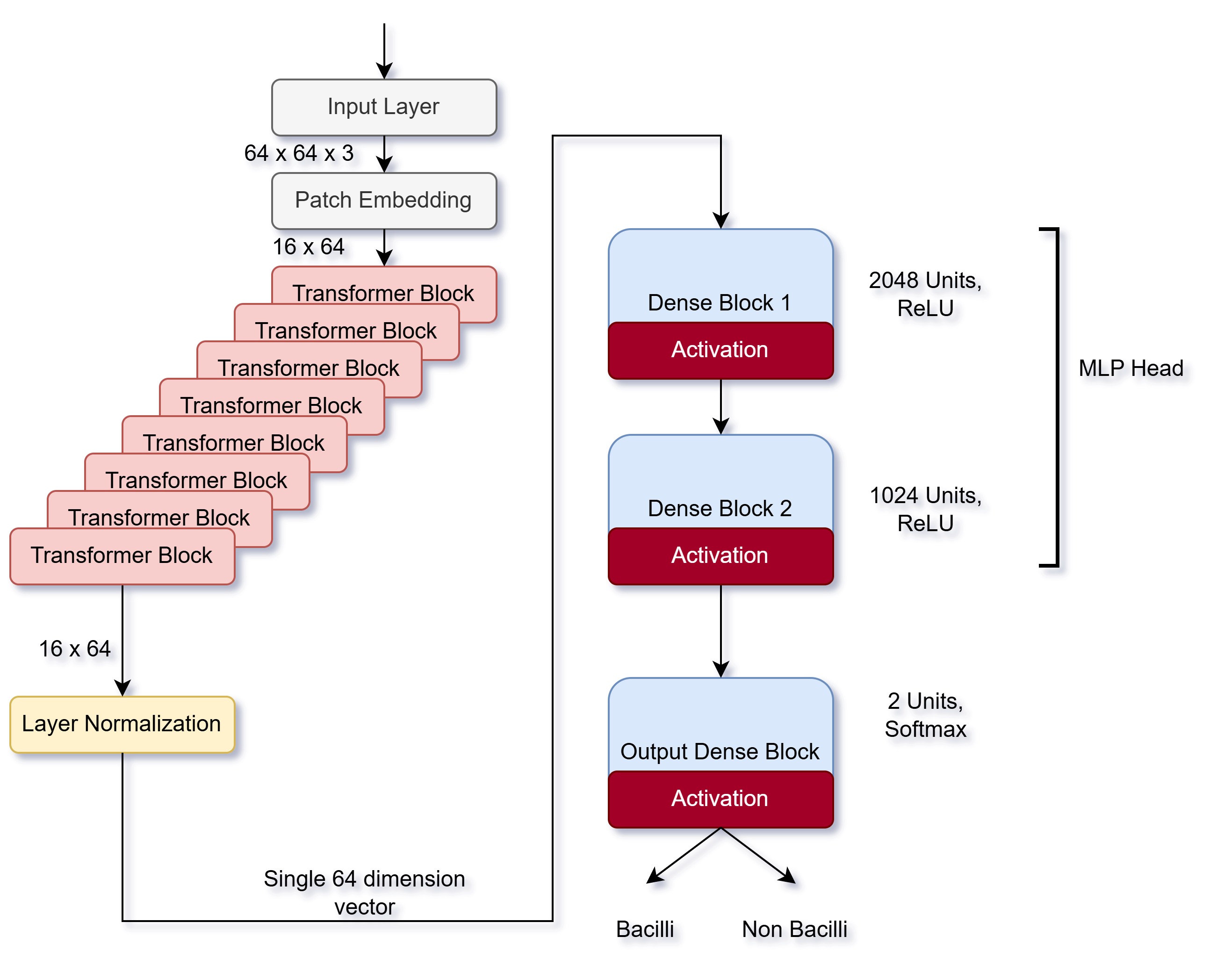}
    \caption{Architecture of Vision Transformer}
    \label{vit}
\end{figure}

Additionally, the TBViT model architecture has been adapted to accommodate smaller patches, enabling the model to efficiently process the ROIs without compromising its ability to capture spatial relationships across the patches. By dividing each ROI into patches and embedding positional information, the model maintains its capacity to learn complex spatial dependencies critical for detecting bacilli. This approach optimizes the ViT for computational efficiency and precision, making it well-suited for binary classification under challenging testing conditions where class imbalance and spatial feature extraction are key factors. The architecture of the Vision Transformer used for classification in the proposed method is shown in Figure \ref{vit}.

\par
Architecturally, TBViT model scales down both image size and patch dimensions to enhance computational efficiency, unlike the original model, which operates on larger images and higher-resolution patches. This choice maintains spatial coherence within patches while enabling the model to operate effectively on smaller datasets and more constrained hardware. Rather than relying on convolutional layers for feature extraction, the model uses self-attention mechanisms across smaller patches to capture both local and global dependencies without the inductive biases inherent to convolutional networks. Despite this downsizing, the model preserves the core ViT principles of patch-based attention and positional embeddings, achieving a flexible architecture that remains effective for binary classification tasks with limited data.

Training is further optimized through a tailored set of callbacks, including early stopping, learning rate reduction, and model checkpointing based on validation performance, which collectively refine the training process and prevent overfitting. In TBViT model, dropout regularization is applied more broadly across transformer layers to enhance generalization on a smaller training set. Model checkpointing ensures that the optimal version of the model is retained, providing stable performance across varying test conditions. These modifications not only make the model more efficient and adaptable for binary classification but also extend the applicability of transformers in image recognition tasks characterized by imbalanced testing conditions.

\section{Experimental Results and Analysis}

The experiments were conducted at the Artificial Intelligence Lab, Department of Computer Applications, Cochin University of Science and Technology, Kerala, India, using a high-performance computing system with advanced specifications. The machine used for training the composite Attention Residual U-Net and TBViT model was equipped with two NVIDIA A100 GPUs, each featuring 40 GB of memory and PCIe 4.0 support, providing a combined total of 6912 CUDA cores and 432 Tensor cores for accelerated deep learning computations. The system was powered by two Intel Xeon Gold 6226R processors, each with 16 cores, clocked at 2.9 GHz, ensuring efficient parallel processing. The memory setup comprised 12 modules of 64 GB DDR4 2933 MHz RDIMM, providing substantial capacity for handling large datasets and complex models. For storage, the system was configured with eight 1.92 TB SATA SSDs, delivering fast data access and high throughput during model training and validation processes.

\par
The experimental workflow was divided into three key phases: the generation of a novel dataset, the assessment of the proposed model's performance on both the newly created dataset and two publicly available datasets, and a comparative analysis to benchmark the model's performance against recent state-of-the-art methods using the same datasets.

\par

The performance of the proposed model for bacilli identification was evaluated across both the segmentation and classification stages. The segmentation performance was assessed using the Jaccard Index and Dice Coefficient to measure the model's effectiveness in accurately delineating bacilli regions. The Jaccard Index is defined as the ratio of the intersection of the predicted segmentation mask and the ground truth mask to their union, providing a robust measure of similarity between the predicted and actual regions in segmentation tasks. The Jaccard Index $J$ is defined as:

\begin{equation}
    J = \frac{A \cap B}{A \cup B}
\end{equation}

where $A \cap B$ represents the number of pixels that are correctly segmented in both the predicted mask (A) and the ground truth mask (B) and $A \cup B$ represents the total number of pixels segmented either in the predicted mask or in the ground truth mask.

Dice coefficient measures the ratio of the intersection of the segmented regions to the average size of the segmented regions in the predicted mask and in the ground truth mask. The Dice Coefficient, D is defined as

\begin{equation}
    D = 2 \times \frac{|A \cap B|}{|A| + |B|}
\end{equation}

Where $|A \cap B|$ represents the number of pixels that are correctly segmented in both the predicted mask (A) and the ground truth mask (B) and $|A|$ and $|B|$ denote the total number of pixels segmented in the predicted mask and ground truth mask, respectively.

\par

The classifier’s performance is evaluated through key metrics, Accuracy, Precision, Recall, and F1 Score, to yield comprehensive insights into the model's effectiveness. Accuracy provides an overall assessment of correctness by calculating the ratio of true positives (TP) and true negatives (TN) to the total instances. However, due to the substantial imbalance in the test set, Accuracy alone may not sufficiently represent the model's capacity to detect bacilli. Precision becomes especially pertinent in this context, as it emphasizes the accuracy of positive predictions by calculating the ratio of true positives to the sum of true positives and false positives (TP + FP), thereby reducing the risk of false positives. Recall, or sensitivity, highlights the model’s capability to capture all relevant cases, expressed as the ratio of true positives to the sum of true positives and false negatives (TP + FN), thus reflecting the model’s effectiveness in identifying bacilli despite class imbalance. The F1 Score combines Precision and Recall into a single metric, providing a balanced assessment that is particularly informative for uneven class distributions. Collectively, these metrics create a rigorous framework for evaluating the classifier’s performance, emphasizing sensitivity and specificity while addressing the challenges posed by imbalanced data.

\par
The Accuracy, Precision, Recall and F1 score are calculated using equations \ref{acc_equ}, \ref{pre}, \ref{re} and \ref{f1} respectively.

\begin{equation}
    Accuracy = \frac{TP + TN}{TP + TN + FP + FN}
    \label{acc_equ}
\end{equation}

\begin{equation}
Precision = \frac{TP}{TP + FP}
\label{pre}
\end{equation}

\begin{equation}
    Recall = \frac{TP}{TP+FN}
    \label{re}
\end{equation}

\begin{equation}
    F1~Score = 2 \times \frac{Precision \times Recall}{Precision + Recall}
    \label{f1}
\end{equation}

\par
Most existing research on bacilli identification from microscopic images does not integrate segmentation into the analysis workflow. Instead, many studies rely on pre-selected ROIs extracted from images, which are then directly processed by a classifier. In contrast, only two recent works \cite{panicker2018automatic} and \cite{mithra2023enhanced} have successfully automated the entire process by incorporating the segmentation stage to identify the ROIs, followed by classification. The performance of the proposed method is benchmarked against these two approaches to validate its efficacy and demonstrate its superiority.

\subsection{DCA-CUSAT Bright Field Microscopic Sputum Smear TB Dataset}
\begin{figure}[htp]
    \centering
    \includegraphics[scale=0.25]{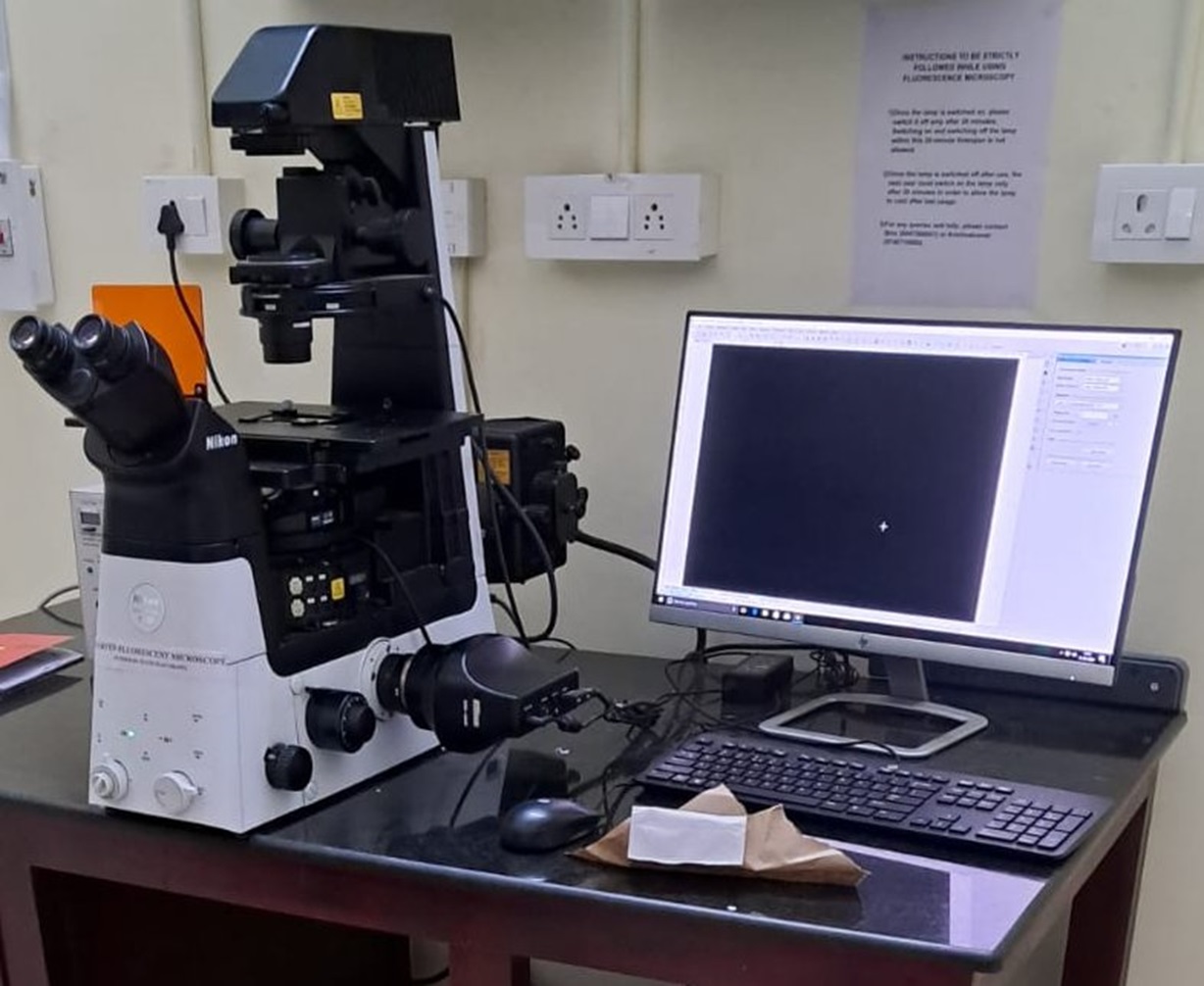}
    \caption{Nikon Ti2-u Eclipse microscope, integrated with the NIS-elements software package at Department of Biotechnology, Cochin University of Science and Technology. }
    \label{mi}
\end{figure}

For this study, we have assembled a database consisting of 101 microscopic images collected from three Ziehl-Neelsen (ZN) stained sputum smear slides, each containing samples from patients diagnosed with Tuberculosis at a severity level of 3+. This database has been designated as the 'DCA-CUSAT Bright Field Microscopic Sputum Smear TB Dataset' (DCA-CUSAT TB dataset). The ZN stained sputum smear slides were sourced from the Government District Tuberculosis Hospital in Ernakulam, Kerala, India, while the imaging was conducted at the Microscopy Facility within the Department of Biotechnology at Cochin University of Science and Technology, Kochi, Kerala, India. The images were captured using the Nikon Ti2-u Eclipse microscope, which was equipped with the NIS-elements software package. Using a Nikon Ti2-u Eclipse microscope, images were captured at 100 × magnification after covering the ZN-stained sputum smear samples with a coverslip and applying a 50\% glycerol solution. This camera integrated system is shown in Figure \ref{mi}.

\par
\begin{figure}[htp]
  \begin{center}
    \subfigure[Slide 1]{\label{sam1}\includegraphics[scale=0.3]{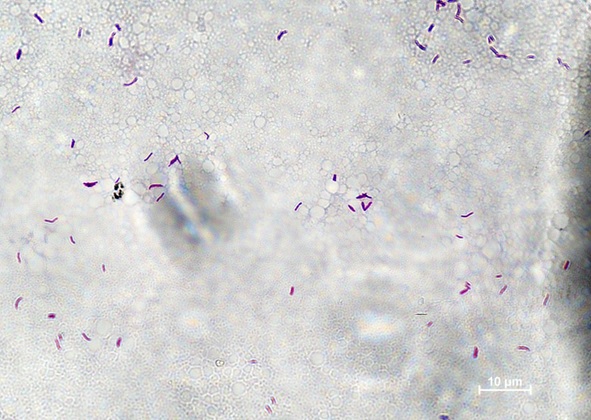}}
    \subfigure[Slide 2]{\label{sam2}\includegraphics[scale=0.3]{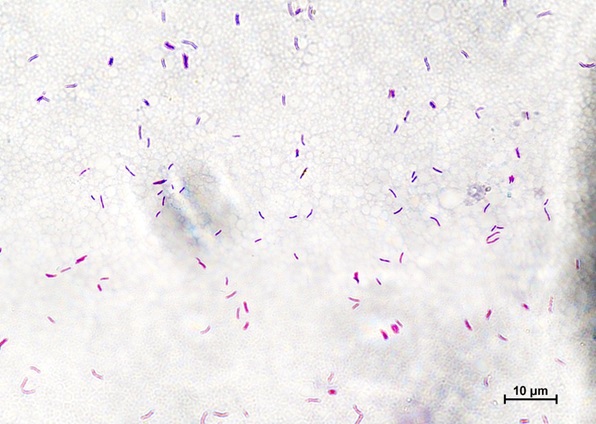}} 
    \subfigure[Slide 3]{\label{sam3}\includegraphics[scale=0.3]{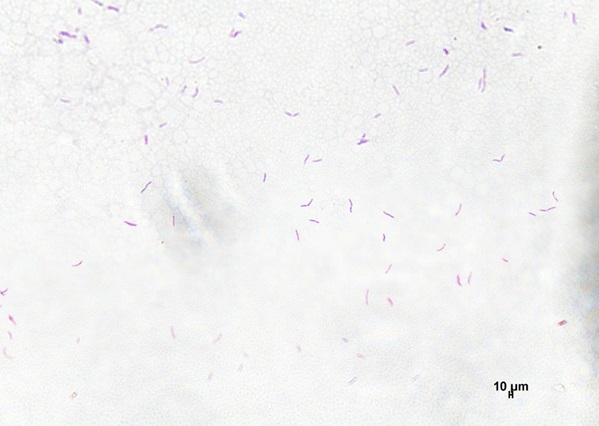}}
  \end{center}
  \caption{Sample images - DCA-CUSAT Bright Field Microscopic Sputum Smear TB Dataset.}
  \label{si}
\end{figure}

A linear pattern that moved methodically in a straight line from top left to bottom right across the specimen produced the images. 31 images were taken from the first slide, and 35 images each from the second and third slides. Every image kept its 2880 $\times$ 2048 pixel resolution. Sample images from each of the three slides are shown in Figure \ref{si}.

\subsection{Performance Analysis of the Proposed Model Using DCA-CUSAT TB dataset}

\par
The DCA-CUSAT TB dataset was used to assess the segmentation performance of the composite Attention Residual U-Net and the classification performance of the TBViT. Of the 101 images, 81 were allocated for training the proposed model, while the remaining 20 were set aside for testing.

\begin{figure}[htp]
  \begin{center}
    \subfigure[Training accuracy]{\label{sam1}\includegraphics[scale=0.7]{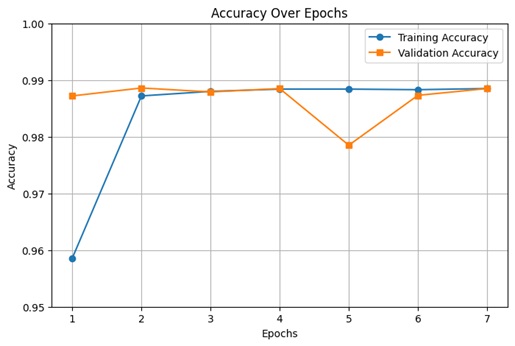}}
    \subfigure[Jaccard Index]{\label{sam2}\includegraphics[scale=0.6]{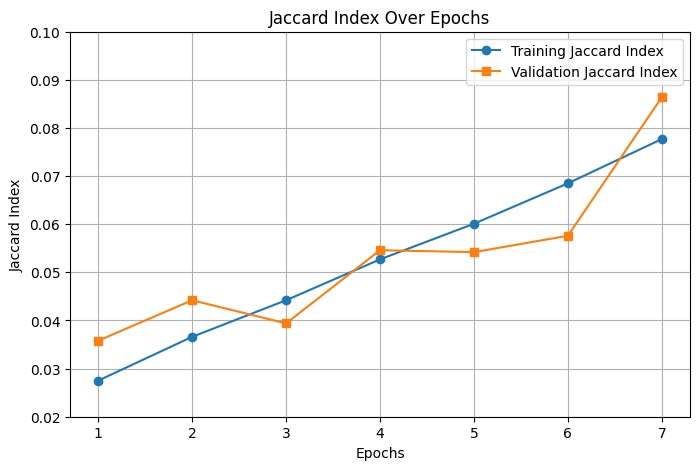}} 
    \subfigure[Loss]{\label{sam3}\includegraphics[scale=0.7]{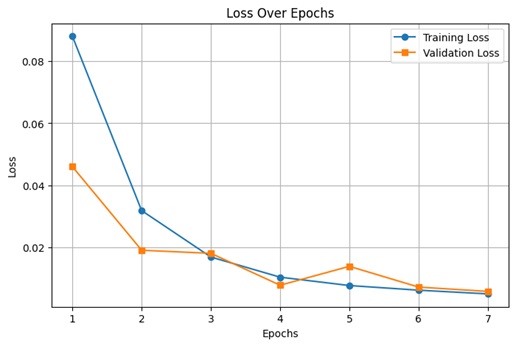}}
  \end{center}
  \caption{The progress of Attention Residual U-Net model training for segmentation on DCA-CUSAT TB dataset.}
  \label{gr}
\end{figure}

\begin{figure}[htp]
  \begin{center}
    \subfigure[Training and Validation accuracy]{\label{sam1}\includegraphics[scale=0.7]{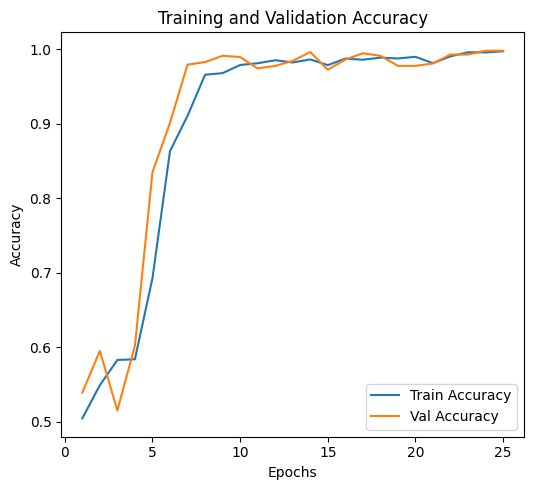}}
    \subfigure[Training and Validation Loss]{\label{sam2}\includegraphics[scale=0.6]{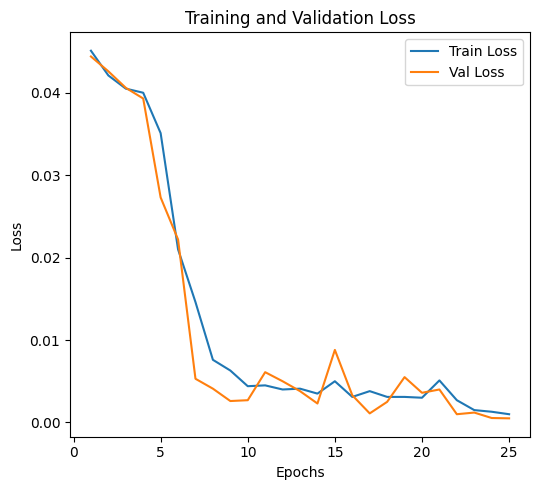}} 
  \end{center}
  \caption{The progress of TBViT model training for classification on DCA-CUSAT TB dataset.}
  \label{aldca}
\end{figure}

\par 
In the segmentation phase, the training images and their corresponding masks were split horizontally and vertically into patches of 256 $\times$ 256 pixels. This approach generated 7128 patches, each accompanied by its binary mask, which were subsequently used to train the Attention Residual U-Net model. Achieving a segmentation accuracy of 98.85\%, the model demonstrated robust performance. Training was scheduled for 20 epochs, with an early stopping criterion applied to mitigate overfitting. However, the training concluded after the $7^{th}$ epoch, as validation accuracy remained stable across consecutive epochs. Improvements in training accuracy, Jaccard Index, and loss values were tracked and are presented in Figure \ref{gr}.

\par
A total of 5886 bacilli regions were extracted from the training images to train the TBViT classifier. To maintain balance and ensure robust learning, an equal number of non-bacilli regions were also extracted from these images. This set of bacilli and non-bacilli regions were then used to form the training dataset, allowing the classifier to distinguish between bacilli and non-bacilli instances effectively. This balanced approach in dataset preparation helps to improve the model's generalization capability and ensures that it performs well in identifying both bacilli and non-bacilli regions during classification. The transformer model was trained over 25 epochs, achieving a training accuracy of 99.26\%. Figure \ref{aldca} illustrates the training accuracy and validation accuracy across 25 epochs, along with the corresponding loss values.

\par
\begin{figure}[htp]
  \begin{center}
    \subfigure[Original RGB test image]{\label{va1}\includegraphics[scale=0.5]{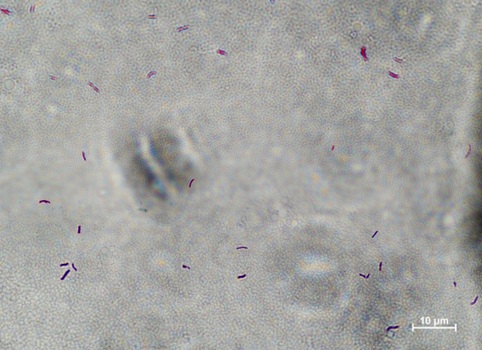}}
    \subfigure[Ground truth mask]{\label{va2}\includegraphics[scale=0.51]{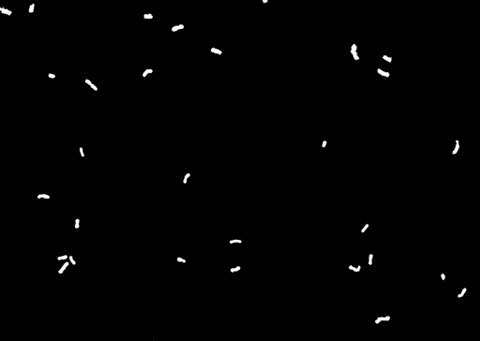}} 
    \subfigure[Segmentation - proposed model]{\label{va3}\includegraphics[scale=0.5]{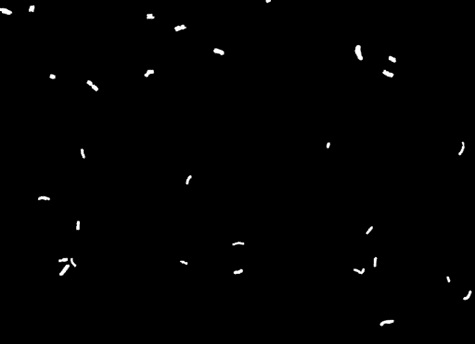}}
    \subfigure[Otsu segmentation described in \cite{panicker2018automatic}]{\label{va4}\includegraphics[scale=0.085]{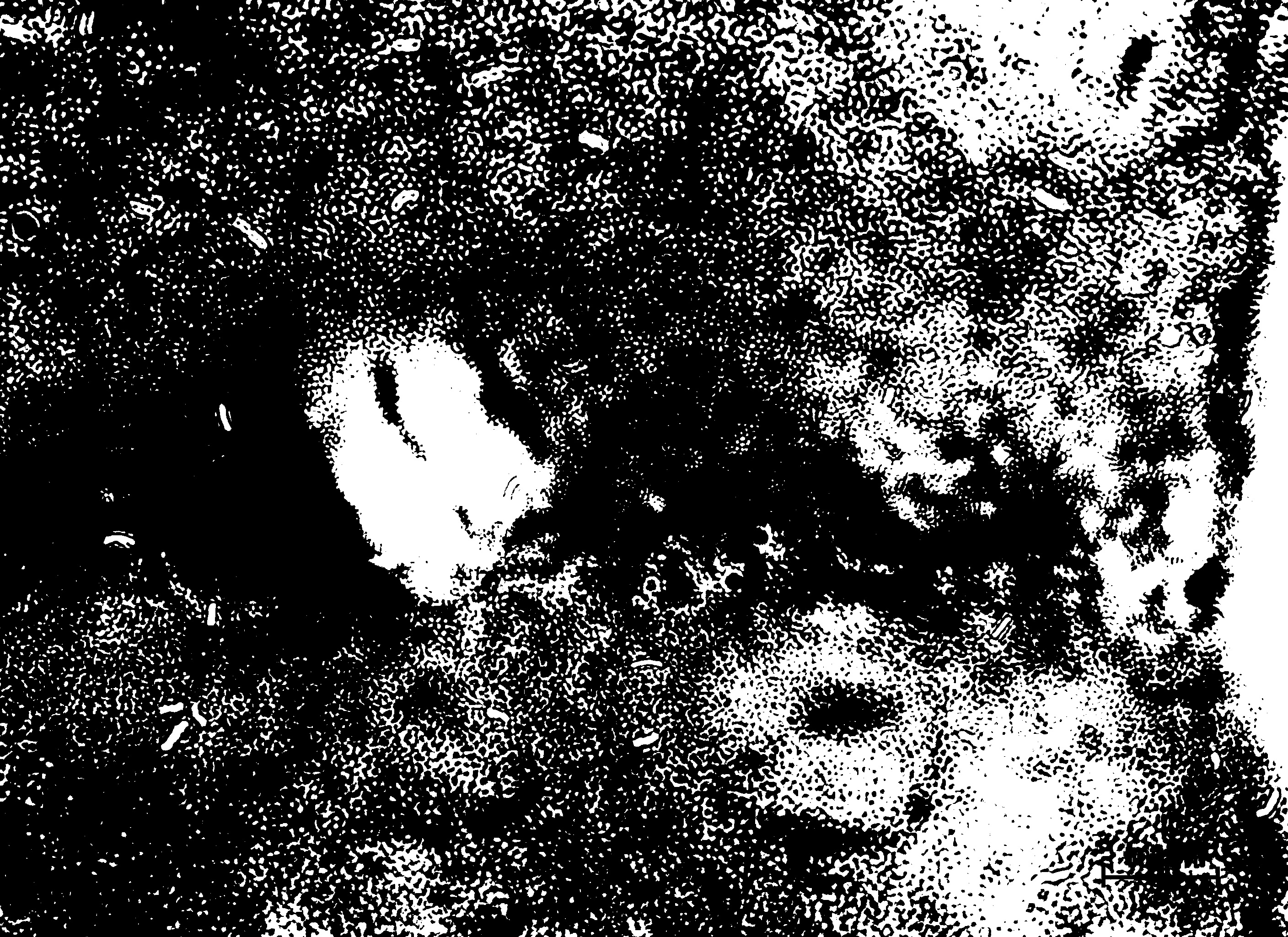}}
    \subfigure[Otsu segmentation described in \cite{mithra2023enhanced}]{\label{va5}\includegraphics[scale=0.5]{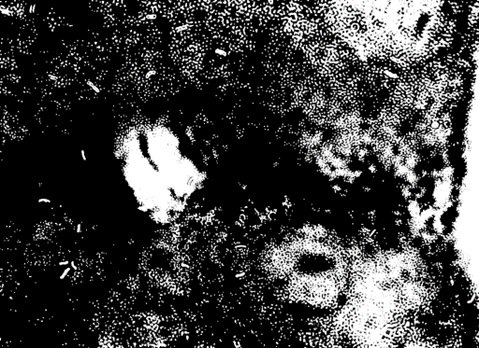}}
  \end{center}
  \caption{Segmentation results of Attention Residual U-Net and Otsu methods described in existing methods}
  \label{va}
\end{figure}

During the testing phase of the segmentation model, a methodology similar to that used during training was employed. The 20 test images were divided into patches of size 256 $\times$ 256, resulting in 1760 patches in total. The selection of the 20 test images was randomized and ensured equal representation across the images from three slides, with 6 images randomly selected from the first slide and 7 from each of the other two slides. These patches were then segmented using the trained composite Attention Residual U-Net to generate predicted masks for each patch. Afterward, the masks from individual patches within an image were combined to create a complete mask for each test image. The segmentation performance of the Attention Residual U-Net, in comparison to the methods described in \cite{panicker2018automatic} and \cite{mithra2023enhanced}, is detailed in Table \ref{seg_per1}. Figure \ref{va} illustrates the segmentation results on a sample test image from the DCA-CUSAT TB dataset. As shown in Table \ref{seg_per1}, the proposed method demonstrates a notable improvement over the segmentation approaches presented in \cite{panicker2018automatic} and \cite{mithra2023enhanced}, a result further corroborated by the visual comparison in Figure \ref{va}.

\begin{table}[htp]
\caption{Performance of segmentation using existing methods and proposed method.}
\label{seg_per1}
\centering
\scalebox{0.7}

\begin{tabular}{|c|c|c|c|c|}
\hline
Dataset                       & \begin{tabular}[c]{@{}c@{}}Evaluation \\ metric\end{tabular} & \multicolumn{1}{l|}{\begin{tabular}[c]{@{}l@{}}Method \\ in \cite{panicker2018automatic}\end{tabular}} & \multicolumn{1}{l|}{\begin{tabular}[c]{@{}l@{}}Method \\ in \cite{mithra2023enhanced}\end{tabular}} & \multicolumn{1}{l|}{\textbf{\begin{tabular}[c]{@{}l@{}}Proposed \\ method\end{tabular}}} \\ \hline
\multirow{2}{*}{DCA-CUSAT TB} & Jaccard Index                                                  & 0.6834                                                                       & 0.7414                                                                       & \textbf{0.9360}                                                                          \\ \cline{2-5} 
                              & Dice coefficient                                               & 0.8119                                                                       & 0.8138                                                                       & \textbf{0.9670}                                                                          \\ \hline
\multirow{2}{*}{Costa}        & Jaccard Index                                                  & 0.4768                                                                       & 0.6361                                                                       & \textbf{0.9845}                                                                          \\ \cline{2-5} 
                              & Dice coefficient                                               & 0.6457                                                                       & 0.7776                                                                       & \textbf{0.9922}                                                                          \\ \hline
\multirow{2}{*}{ZNSM-iDB}     & Jaccard Index                                                  & 0.7328                                                                       & 0.7715                                                                       & \textbf{0.9767}                                                                          \\ \cline{2-5} 
                              & Dice coefficient                                               & 0.8457                                                                       & 0.8710                                                                       & \textbf{0.9882}                                                                          \\ \hline
\end{tabular}
\end{table}

\par 
To assess the performance of the proposed bacilli detection method following segmentation, contour analysis was carried out to identify suspected bacilli regions on each mask corresponding to the test images. The regions whose area is above a threshold of 200, were extracted from the images and treated as Regions of Interest (ROIs). The ROIs were then classified as either bacilli or non-bacilli using the proposed TBViT classifier. The performance of the suggested method was compared against existing methods \cite{panicker2018automatic} and \cite{mithra2023enhanced} on the DCA-CUSAT TB dataset and the comparative results are provided in Table \ref{pcr_ec}. The quantitative metrics, including accuracy, precision, recall, and F1 score, as detailed in Table \ref{pcr_ec}, demonstrate that the proposed model exceeds other methods in bacilli identification.

\begin{table}[htp]
\centering
\caption{Performance comparison of existing methods and proposed method on the DCA-CUSAT TB dataset}
\label{pcr_ec}
\begin{tabular}{|c|c|c|c|c|}
\hline
Method                                                                                  & Accuracy & Precision & Recall & F1-Score \\ \hline
\begin{tabular}[c]{@{}c@{}}Method in \cite{panicker2018automatic}\end{tabular} & 0.6773   & 0.6711    & 0.6    & 0.6335   \\ \hline
\begin{tabular}[c]{@{}c@{}}Method in \cite{mithra2023enhanced}\end{tabular}                & 0.7119   & 0.9       & 0.75   & 0.8181     \\ \hline
\begin{tabular}[c]{@{}c@{}}\textbf{Proposed} \textbf{Method}\end{tabular}                              & \textbf{0.9941}     &\textbf{ 1 }      &\textbf{ 0.9939}   & \textbf{0.9969  }   \\ \hline
\end{tabular}
\end{table}

\subsection{Performance Analysis of the Proposed Model Using Costa Dataset}
The proposed model was assessed and compared with other existing methods using the Costa dataset.\cite{ya12-j913-22}. A total of 90 images were randomly selected from this dataset, encompassing both high and low background densities. The selected images also varied in bacilli density, with some containing a high concentration of bacilli while others had lower densities. Of these, 72 images were used during the training phase of the model, while the remaining 18 images were set aside for testing. All images in the dataset had a resolution of 2816 $\times$ 2048.
\par
\begin{figure}[htp]
  \begin{center}
    \subfigure[Training accuracy]{\label{sam1}\includegraphics[scale=0.5]{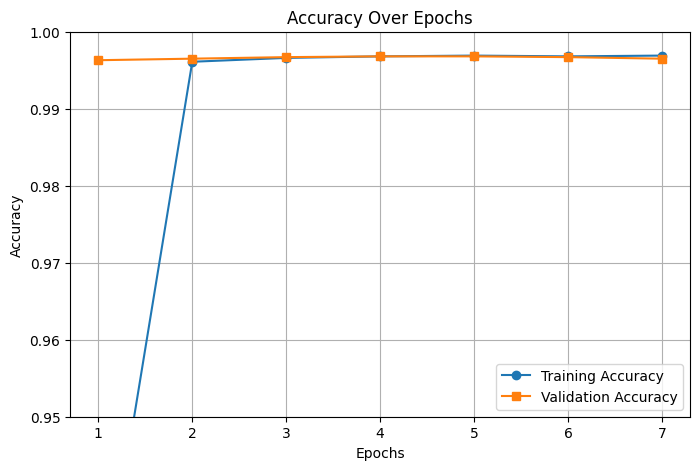}}
    \subfigure[Jaccard Index]{\label{sam2}\includegraphics[scale=0.5]{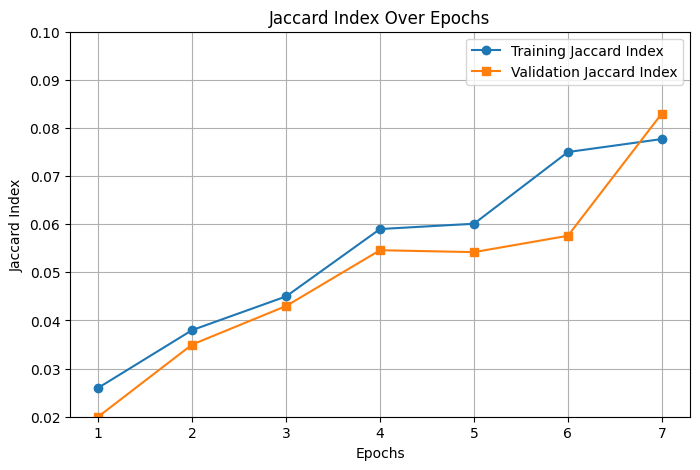}} 
    \subfigure[Loss]{\label{sam3}\includegraphics[scale=0.5]{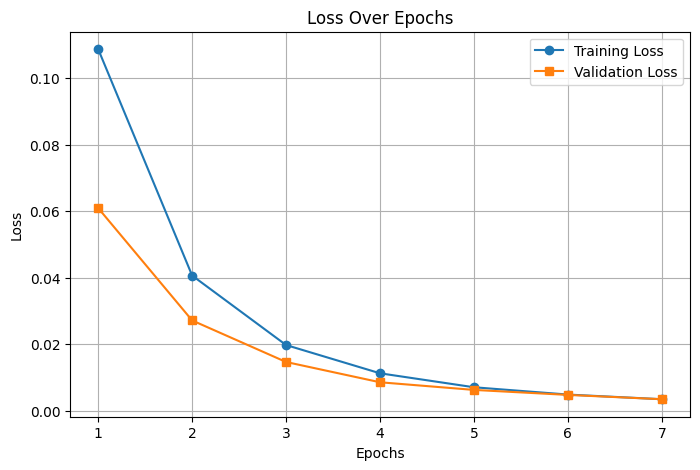}}
  \end{center}
  \caption{The progress of the Attention Residual U-Net training on Costa dataset.}
  \label{gr3}
\end{figure}

During the training phase, the segmentation model was trained on a comprehensive dataset obtained from 72 images. From these, 6336 patches, each sized 256 $\times$ 256, were extracted. These patches, paired with their corresponding ground truth masks, were used to train the proposed composite Attention Residual U-Net model, which demonstrated excellent performance. The model achieved a high training accuracy of 99.65\%. Figure \ref{gr3} provides a visual representation of the accuracy progression, along with the Jaccard index and loss metrics.

\par
\begin{figure}[htp]
  \begin{center}
    \subfigure[Training and Validation accuracy]{\label{sam1}\includegraphics[scale=0.5]{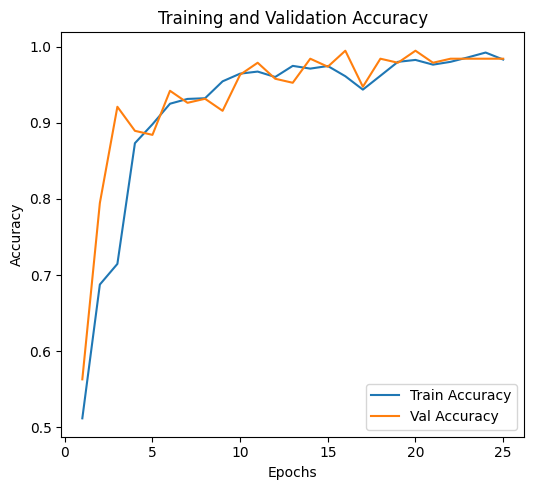}}
    \subfigure[Training and Validation Loss]{\label{sam2}\includegraphics[scale=0.5]{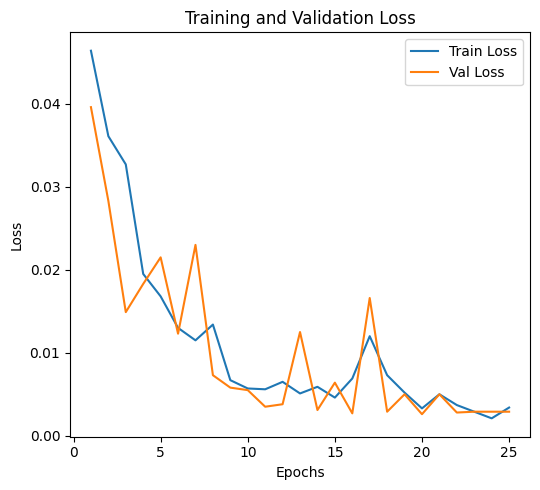}} 
  \end{center}
    \caption{The progress of TBViT model training for classification on Costa dataset.}
        \label{alcosta}
\end{figure}

 
For training the TBViT classifier on the Costa dataset, a balanced set of 1909 bacilli-containing segments and 1909 non-bacilli segments were carefully extracted from the training images. This ensured an equal representation of both classes, allowing the model to effectively learn to distinguish between bacilli and non-bacilli regions. The training process was carried out over 25 epochs, during which the model was fine-tuned to optimize its performance. After the training phase, the TBViT classifier achieved an accuracy of 97\%, demonstrating its effectiveness in classifying bacilli within the dataset.Figure \ref{alcosta} presents the training and validation accuracy over 25 epochs, along with the associated loss values.

\par
During the testing phase, the trained composite Attention Residual U-Net successfully segmented 1584 patches obtained from 18 test images. These segmented patches were subsequently combined to reconstruct the masks corresponding to each test image. The performance of the proposed segmentation method, along with a comparison to the methods in \cite{panicker2018automatic} and \cite{mithra2023enhanced}, was evaluated on the Costa dataset. The results of this comparative analysis are presented in Table \ref{seg_per1}.

\par 
The ROIs obtained from the original images through the proposed segmentation model and contour analysis are classified using the trained TBViT model. Table \ref{per_en_ex} presents a performance comparison between the proposed method and existing approaches on the Costa dataset. The proposed model's superiority over other methods is clearly demonstrated in Table 5, highlighting its outstanding performance on the Costa dataset as well.

\begin{table}[htp]
\centering
\caption{Performance comparison of existing methods and proposed method on the Costa dataset}
\label{per_en_ex}
\begin{tabular}{|c|c|c|c|c|}
\hline
\textbf{Method}                                                                                            & \textbf{Accuracy} & \textbf{Precision} & \textbf{Recall} & \textbf{F1-Score} \\ \hline
\begin{tabular}[c]{@{}c@{}}Method  in \cite{panicker2018automatic}\end{tabular} & 0.6               & 0.67               & 0.6             & 0.6330              \\ \hline
\begin{tabular}[c]{@{}c@{}}Method in \cite{mithra2023enhanced}\end{tabular}    & 0.73              & 0.8                & 0.758           & 0.7784              \\ \hline
\textbf{\begin{tabular}[c]{@{}c@{}}Proposed Method\end{tabular}}                                        & \textbf{0.9904}     & \textbf{0.9966}      & \textbf{0.9933}   & \textbf{0.9949}     \\ \hline
\end{tabular}
\end{table}

\subsection{Performance Analysis of the Proposed Model Using ZNSM-iDB Dataset}
To evaluate performance, the proposed method was tested on the ZNSM-iDB \cite{shah2017ziehl} dataset, which consists of images with a variety of resolutions and backgrounds in different colors. A total of 90 images, each with dimensions of 2592 $\times$ 1944, were randomly selected from the dataset. Out of these, 72 images were used for training, and the remaining 18 were reserved for testing.

\par
\begin{figure}[htp]
  \begin{center}
    \subfigure[Training accuracy]{\label{sam1}\includegraphics[scale=0.5]{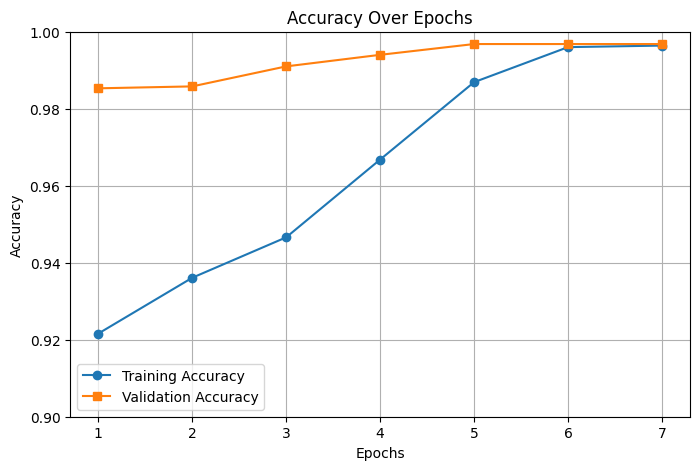}}
    \subfigure[Jaccard Index]{\label{sam2}\includegraphics[scale=0.5]{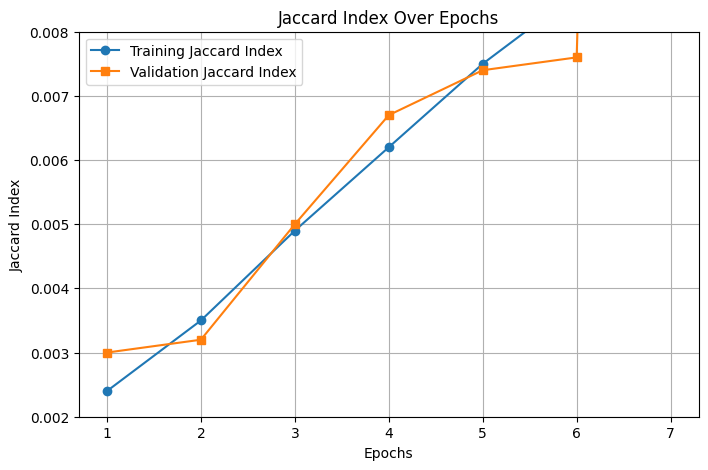}} 
    \subfigure[Loss]{\label{sam3}\includegraphics[scale=0.5]{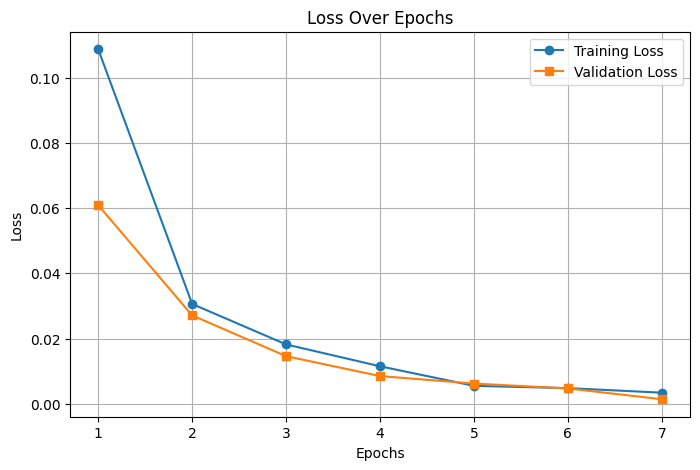}}
  \end{center}
  \caption{The progress of Attention Residual U-Net training on ZNSM-iDB dataset.}
  \label{gr2}
\end{figure}

During the training phase of the segmentation model, 5040 patches, each measuring 256 $\times$ 256, were extracted from 72 images. These patches, along with their respective masks, were used to train the composite Attention Residual U-Net model, resulting in an accuracy of 99.68\%. The accuracy, Jaccard index, and loss values recorded during training are illustrated in Figure \ref{gr2}.

\par
\begin{figure}[htp]
  \begin{center}
    \subfigure[Training and Validation accuracy]{\label{sam1}\includegraphics[scale=0.5]{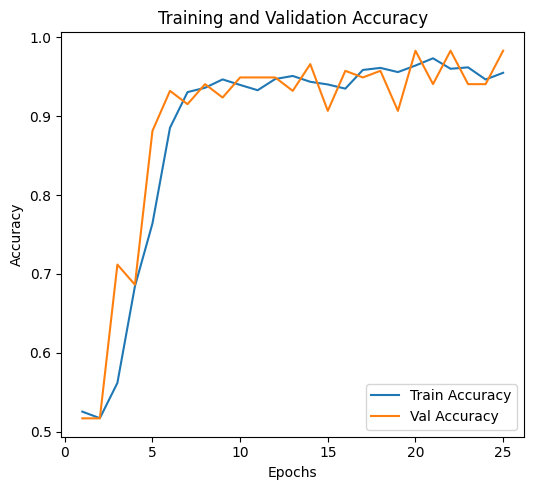}}
    \subfigure[Training and Validation Loss]{\label{sam2}\includegraphics[scale=0.5]{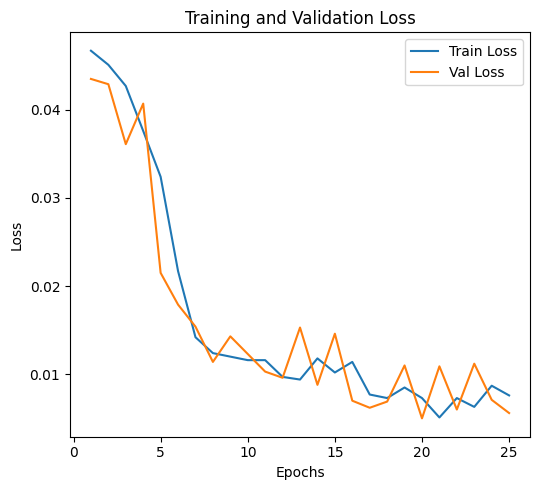}} 
    \end{center}
     \caption{The progress of TBViT model training for classification on ZNSM-iDB dataset.}
        \label{alzndb}
\end{figure}

Both bacilli and non-bacilli regions were extracted from the dataset for training the custom Vision Transformer, TBViT. Specifically, 1184 bacilli segments and 1184 non-bacilli segments were obtained from the training images. The TBViT model was trained over 25 epochs, resulting in a final accuracy of 97.5\%.Training and validation accuracy across 25 epochs, along with the corresponding loss values, are illustrated in Figure \ref{alzndb}.

\par
Following the training process, 1260 test patches were segmented using the Attention Residual U-Net model. Each patch, with dimensions of 256 $\times$ 256, was generated by dividing the 18 test images from the ZNSM-iDB dataset. The performance of the proposed segmentation model in terms of Jaccard Index and Dice coefficient, along with the existing methods discussed in \cite{panicker2018automatic} and \cite{mithra2023enhanced}, are presented in Table \ref{seg_per1}.

\par
Finally, tuberculosis bacilli are detected in each test image using the proposed approach. The effectiveness of this method on the ZNSM-iDB dataset is evaluated and benchmarked against the techniques outlined in \cite{panicker2018automatic} and \cite{mithra2023enhanced}. The comparative outcomes, shown in Table \ref{znaccuracy}, clearly demonstrate that the proposed method outperforms existing techniques in terms of accuracy, precision, recall, and F1 score.

\begin{table}[htp]
\centering
\caption{Performance of various identification methods on the ZNSM-iDB dataset.}
\label{znaccuracy}
\begin{tabular}{|c|c|c|c|c|}
\hline
\textbf{Method}                                                                                            & \textbf{Accuracy} & \textbf{Precision} & \textbf{Recall} & \textbf{F1-Score} \\ \hline
\begin{tabular}[c]{@{}c@{}}Method in \cite{panicker2018automatic}\end{tabular} & 0.63              & 0.6                & 0.66            & 0.6285              \\ \hline
\begin{tabular}[c]{@{}c@{}}Method in \cite{mithra2023enhanced}\end{tabular}    & 0.74              & 0.7                & 0.765           & 0.7310             \\ \hline
\textbf{\begin{tabular}[c]{@{}c@{}}Proposed Method\end{tabular}}                                        & \textbf{0.9601}     & \textbf{0.9892}      & \textbf{0.9435}   & \textbf{0.9659}     \\ \hline
\end{tabular}
\end{table}

\par
In \cite{panicker2018automatic} and \cite{mithra2023enhanced}, two modified versions of the Otsu algorithm were employed for image binarization, offering a more computationally efficient solution compared to the proposed segmentation model. However, these methods resulted in noticeably lower Jaccard Index and dice coefficient scores, as detailed in Table \ref{seg_per1}, denoting their subpar segmentation performance. Additionally, the a visual analysis using Figure \ref{va} reveals that the segmentation produced by our model outclasses that of the modified Otsu techniques in \cite{panicker2018automatic} and \cite{mithra2023enhanced}, closely aligning with the ground truth masks.

\par
The proposed Vision Transformer model TBViT, delivered outstanding results in classification, outperforming the approaches in \cite{panicker2018automatic} and \cite{mithra2023enhanced}. Unlike traditional methods that rely on convolutional layers, vision transformer employs self-attention mechanisms to analyze image patches, allowing it to capture complex spatial relationships more effectively. This ability to handle global context makes it highly suited for classification tasks, where it excels in terms of accuracy and robustness. In contrast to earlier methods, which may struggle with intricate patterns, the presented TBViT architecture provides a clear advantage, leading to superior performance across various metrics.

\section{Conclusion}
The proposed methodology, which integrates a composite Attention Residual U-Net for segmentation and a custom Vision Transformer, TBViT for classification, offers a significant advancement in tuberculosis detection from bright-field microscopic sputum smear images. The level of automation achieved is notably higher than that of existing methods, and the segmentation performance outpaces previous benchmarks. Additionally, the curation of the new 'DCA-CUSAT Bright Field Microscopic Sputum Smear TB Dataset' enhances the robustness of the experiments. Overall, this approach not only improves diagnostic accuracy but also streamlines the detection process, making it a valuable tool for effective tuberculosis treatment strategies. Future enhancements could focus on expanding the dataset and developing techniques to accurately count the number of bacilli within clusters to improve diagnostic precision.

\subsection*{Acknowledgements}
We extend our sincere thanks to Dr. Sarath G Rao, District TB officer and technical staff of District TB Hospital, Ernakulam, Kerala, India for  providing ZN-stained sputum smear samples necessary for capturing microscopic images, and providing ground truth information which were crucial for conducting our research.


%
%

\end{document}